\documentclass[sigconf,10pt]{acmart}

\usepackage{amsmath}
\usepackage{booktabs}
\usepackage{float}
\usepackage{graphicx}
\usepackage{balance}
\balance
\usepackage[inline]{enumitem}

\AtBeginDocument{%
  }

\copyrightyear{2026}
\acmYear{2026}
\setcopyright{cc}
\setcctype{by}
\acmConference[HumanSys '26]{The 4th International Workshop on Human-Centered Sensing, Modeling, and Intelligent Systems}{October 26--30, 2026}{Austin, TX, USA}
\acmBooktitle{The 4th International Workshop on Human-Centered Sensing, Modeling, and Intelligent Systems (HumanSys '26), October 26--30, 2026, Austin, TX, USA}
\acmDOI{10.1145/3842436.3843818}
\acmISBN{979-8-4007-2959-1/2026/10}

\begin{document}

\title{RAST: Resolution-Aware Privileged Structure Transfer for Low-Resolution Audio Activity Recognition}

\author{Ji Hwan Park}
\affiliation{%
  \institution{The University of Texas at Austin}
  \city{Austin}
  \state{Texas}
  \country{USA}}
\email{jihwanpark98@utexas.edu}

\author{Gautham Krishna Gudur}
\affiliation{%
  \institution{The University of Texas at Austin}
  \city{Austin}
  \state{Texas}
  \country{USA}}
\email{gauthamkrishna@utexas.edu}

\author{Yufei Shen}
\affiliation{%
  \institution{The University of Texas at Austin}
  \city{Austin}
  \state{Texas}
  \country{USA}}
\email{shenyufei@utexas.edu}

\author{Dawei Liang}
\affiliation{%
  \institution{The University of Texas at Austin}
  \city{Austin}
  \state{Texas}
  \country{USA}}
\email{dawei.liang@utexas.edu}

\author{Edison Thomaz}
\affiliation{%
  \institution{The University of Texas at Austin}
  \city{Austin}
  \state{Texas}
  \country{USA}}
\email{ethomaz@utexas.edu}

\renewcommand{\shortauthors}{Park et al.}

\begin{abstract}
Audio is increasingly used for human activity recognition (HAR) because it 
captures object interactions, environmental events, and contextual cues in 
everyday environments. High-resolution (HR) audio provides rich acoustic 
information for model development but incurs substantial energy and storage 
costs and may expose sensitive speech content. Low-resolution (LR) audio 
offers a more privacy-preserving and resource-efficient alternative for
deployment, but reduced sampling rates can remove acoustic cues essential for 
activity recognition, leading to significant performance degradation. 
We formulate this training–deployment mismatch as 
\emph{sensor-resolution privileged learning}, in which HR audio is available 
during training, while inference relies exclusively on LR audio. We propose 
\emph{RAST}, a resolution-aware transfer framework that compresses HR teacher 
representations by preserving token-level information and neighborhood 
structure before performing localized HR--LR alignment. Experiments on the 
SAMoSA and AudioIMU datasets show that RAST consistently outperforms LR-only 
training and direct teacher-transfer baselines, improving LR-only recognition 
by up to $\sim7.8\%$ while requiring only LR audio at inference.\looseness=-1
\end{abstract}

\ccsdesc[500]{Human-centered computing}
\ccsdesc[500]{Computing methodologies~Machine learning}
\ccsdesc[500]{Computing methodologies~Learning paradigms}

\keywords{human-centered sensing, audio sensing, privileged learning}

\maketitle

\section{Introduction}

Audio is an attractive modality for human-centered sensing because everyday 
sounds encode rich contextual information about human behavior, object 
interactions, environmental context, and so on. These capabilities have 
motivated acoustic sensing for activities of daily living, classification from low-sampled microphone input, and wearable HAR \cite{pires2017acousticadl,liang2019audioadl,liang2024lowsampled}.
However, continuous acoustic sensing imposes practical and social constraints,
including substantial bandwidth, storage, computation, and energy costs
\cite{georgiev2014dspear}, while always-on microphones raise concerns about
privacy, trust, and unintended speech capture
\cite{chhaglani2025featuresense}. Recent work has therefore explored
restricted-bandwidth and low-resolution audio as privacy-aware sensing
alternatives, including analyses of speech intelligibility under bandwidth
reduction \cite{liu2024lowfrequencyprivacy}. Audio-based HAR thus presents a
broader human-centered systems challenge: reliably recognizing meaningful behavior 
while limiting sensing cost, sensitive-data exposure, and resource consumption
\cite{lu2009soundsense,consolvo2008activity}.\looseness=-1

\begin{figure}[t]
\centering
\includegraphics[width=0.95\columnwidth]{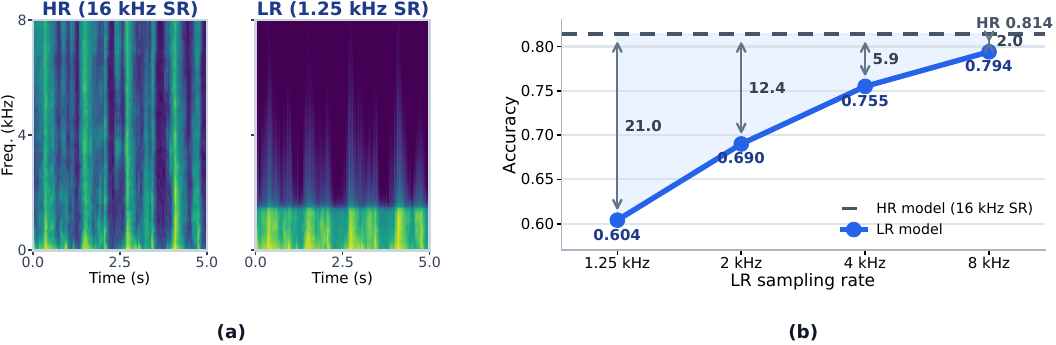}
\Description{Resolution gap between High-resolution (HR) and Low-resolution 
(LR) in audio-based activity recognition.}
\caption{Resolution gap in audio-based activity recognition. (a) High-resolution
(HR) and Low-resolution (LR) views of the same audio window. (b) LR accuracy
decreases as the sampling rate is reduced relative to the 16-kHz HR reference.
\looseness=-1}
\label{fig:resolution_gap}
\end{figure}

Low-resolution (LR) audio is well suited for resource-constrained and
privacy-sensitive deployment, but reducing the sampling rate can remove
acoustic cues critical for distinguishing activities.
High-resolution (HR) audio preserves high-frequency and transient structure 
associated with activities such as chopping, brushing, clapping,
pouring, and appliance use whereas a deployed model may observe only the 
LR signal. Figure~\ref{fig:resolution_gap} illustrates this resolution gap: 
LR audio suppresses salient acoustic structure, and held-out recognition 
accuracy decreases as the sampling rate diverges from the 16-kHz HR 
reference. Prior work has shown that large-scale pretrained audio 
representations can improve downstream sound and activity recognition
\cite{aytar2016soundnet,kong2020panns,gudurpcl26}, while recent work has
explored self-distillation for low-sampled microphone input
\cite{liang2024lowsampled}. Motivated by this gap, we study a
\emph{privileged-learning} setting in which HR audio is available only 
during training, while inference operates exclusively on LR audio.
The objective is to transfer discriminative HR information without 
requiring HR sensing at deployment.\looseness=-1

This setting extends beyond conventional teacher-student transfer.
Classical knowledge distillation typically compresses a larger or more capable
teacher into a smaller student given comparable inputs
\cite{hinton2015distilling}. Generalized distillation and learning using privileged information allow access 
to additional training-only information 
\cite{vapnik2009new,lopezpaz2016unifying}, while cross-modal 
distillation transfers supervision across paired modalities
\cite{gupta2016crossmodal,aytar2016soundnet}. In our setting, the primary 
challenge is instead \emph{limited input observability}: the LR student receives 
a systematically degraded view of the same acoustic event. Consequently, some 
evidence encoded by the HR teacher may be absent from the LR signal, making 
direct imitation of the full HR representation both restrictive 
and potentially infeasible.\looseness=-1

To address this challenge, we propose \emph{Resolution-Aware Privileged Structure
Transfer (RAST)}, a structured HR-to-LR transfer framework tailored to 
sensor-resolution mismatch. Rather than forcing the LR student to reproduce the 
complete HR embedding space, RAST compresses teacher representations into 
intermediate structure comprising HR-derived tokens that summarize coarse teacher 
states and teacher neighborhoods that encode local geometry. The LR student is 
first warm-started using this privileged structure and then refined through local 
hard-negative-mining (HNM) alignment. This design builds on prior work showing 
that intermediate hints, relational geometry, contrastive objectives, and 
information bottlenecks can transfer knowledge beyond final logits
\cite{romero2015fitnets,park2019rkd,tian2020crd,tishby2000information}. RAST 
therefore focuses transfer on HR-derived structure that is both discriminative and
recoverable from LR audio, avoiding imitation of fine-grained information
that may be unobservable after resolution reduction.\looseness=-1

This paper makes three contributions. 
\begin{enumerate*}[label=\textbf{(\arabic*)}]
\item We formulate lower-resolution audio HAR as a \emph{sensor-resolution 
privileged learning} problem, where an LR student must learn from an HR 
teacher despite limited input observability.
\item We design \emph{RAST}, a resolution-aware HR-to-LR transfer
framework that combines HR-derived token bottlenecks, teacher-neighborhood 
initialization, and local representation alignment.
\item We evaluate RAST on two audio-HAR datasets – SAMoSA \cite{mollyn2022samosa} 
and AudioIMU \cite{liang2022audioimu}, under subject-independent protocols. 
Relative to LR-only training, RAST improves balanced accuracy and macro-F1 
by $6.7\%$ and $7.5\%$ on SAMoSA, and by $6.1\%$ and $7.8\%$ on AudioIMU 
respectively.\looseness=-1
\end{enumerate*}

Together, these results demonstrate that privileged HR audio available only 
during training can substantially improve LR activity recognition while 
preserving an exclusively low-resolution sensing path at inference.\looseness=-1

\section{Related Work}

\paragraph{Knowledge transfer with privileged information.}
Knowledge distillation transfers teacher knowledge through softened outputs \cite{hinton2015distilling}, with subsequent work extending transfer to 
intermediate features, contrastive objectives, and relational structure \cite{romero2015fitnets,tian2020crd,park2019rkd}. Learning using privileged 
information allows additional training-only information \cite{vapnik2009new}, 
while generalized distillation integrates such information into teacher--student 
transfer \cite{lopezpaz2016unifying}. Related paradigms include 
dataset distillation for compact audio-classification training sets 
\cite{gudur_enlsp24}, zero-shot distillation for statistical heterogeneity 
in privacy-preserving federated learning \cite{gudur2021zeroshotfl}, and 
cross-modal distillation across paired modalities 
\cite{gupta2016crossmodal,aytar2016soundnet}. HR-to-LR audio transfer introduces
a distinct observability constraint because the student receives a lower-resolution
view of the same event. Directly matching HR predictions or representations may
therefore impose ill-suited targets that are partially unobservable or unrecoverable 
from LR input.\looseness=-1

\paragraph{Deployment-aware activity sensing.}
Human-centered sensing must balance recognition accuracy with sensing cost,
privacy, participant burden, and robustness. Mobile acoustic sensing has 
enabled context and activity recognition \cite{liang2019audioadl}, while other 
prior work has characterized deployment constraints, user variability, and 
sensor heterogeneity
\cite{consolvo2008activity,wang2019deepactivitysurvey,stisen2015smart,sundaramoorthy_emdl18}.
Privacy and trust concerns around always-on audio have further motivated 
low-resolution and bandwidth-restricted sensing \cite{chhaglani2025featuresense}, 
including self-distillation for 1--2 kHz microphone inputs \cite{liang2024lowsampled} 
and discussions on privacy--utility tradeoffs in low-frequency speech 
\cite{liu2024lowfrequencyprivacy}. RAST leverages HR audio only as privileged training 
supervision while requiring only LR audio at inference.\looseness=-1

\paragraph{Compact teacher structure.}
Information-bottleneck methods seek compact representations that preserve 
task-relevant information while suppressing unnecessary detail
\cite{tishby2000information,alemi2017deepvib}, while discrete representation 
learning provides mechanisms for constructing compact latent codes 
\cite{oord2017vqvae}. More broadly, hint-based, relational, and contrastive 
distillation demonstrate that intermediate structure can convey useful teacher 
knowledge beyond hard labels or final logits 
\cite{romero2015fitnets,park2019rkd,tian2020crd}. RAST adapts this principle 
to sensor-resolution mismatch by transferring compact HR-derived tokens and 
local teacher-neighborhood structure instead of requiring the LR student to 
reproduce the full HR embedding space.\looseness=-1
\begin{table*}[t]
\centering
\caption{Subject-held-out audio activity recognition performance. BA denotes
balanced accuracy. The HR teacher is shown as the high-resolution reference,
while the remaining methods use LR input at inference. Parentheses report
relative gains over the LR-only baseline.\looseness=-1}
\label{tab:main_results}
\begin{tabular}{lcccc}
\toprule
& \multicolumn{2}{c}{SAMoSA} & \multicolumn{2}{c}{AudioIMU} \\
\cmidrule(lr){2-3}\cmidrule(lr){4-5}
Method & BA & Macro-F1 & BA & Macro-F1 \\
\midrule
HR teacher    & $0.8352 \pm 0.0083$                                      & $0.8307 \pm 0.0075$                                      & $0.7246 \pm 0.0017$                                      & $0.7066 \pm 0.0032$                                      \\
LR-only       & $0.6123 \pm 0.0054$                                      & $0.6030 \pm 0.0046$                                      & $0.5157 \pm 0.0033$                                      & $0.4933 \pm 0.0031$                                      \\
KD            & $0.6109 \pm 0.0035$                                      & $0.6053 \pm 0.0047$                                      & $0.5124 \pm 0.0054$                                      & $0.4843 \pm 0.0064$                                      \\
InfoNCE       & $0.6137 \pm 0.0041$                                      & $0.6086 \pm 0.0016$                                      & $0.5161 \pm 0.0038$                                      & $0.4994 \pm 0.0040$                                      \\
ProtoNCE      & $0.6112 \pm 0.0065$                                      & $0.6055 \pm 0.0079$                                      & $0.5154 \pm 0.0023$                                      & $0.4921 \pm 0.0036$                                      \\
HNM           & $0.6218 \pm 0.0082$                                      & $0.6093 \pm 0.0086$                                      & $0.5130 \pm 0.0019$                                      & $0.5079 \pm 0.0032$                                      \\
\textbf{RAST} & $\mathbf{0.6532 \pm 0.0072}^{\scriptscriptstyle(+6.7\%)}$    & $\mathbf{0.6482 \pm 0.0067}^{\scriptscriptstyle(+7.5\%)}$    & $\mathbf{0.5474 \pm 0.0042}^{\scriptscriptstyle(+6.1\%)}$    & $\mathbf{0.5320 \pm 0.0055}^{\scriptscriptstyle(+7.8\%)}$    \\
\bottomrule
\end{tabular}
\end{table*}

\section{Methodology}

\subsection{Problem Formulation}

Let $\mathcal{D}=\{(x_i^H,x_i^L,y_i)\}_{i=1}^N$ denote the training set, where
$x_i^H$ and $x_i^L$ are paired high-resolution (HR) and low-resolution (LR)
views of the same audio segment, respectively, and $y_i \in \{1,\ldots,C\}$
is the activity label. The HR view is available only during training.
At inference time, the deployed model observes only $x^L$ and
must predict its activity label. This formulation separates the deployment
constraint from the model-development setting: richer HR audio can be collected
or retained for training, while the deployed classifier remains an LR-only system.
\looseness=-1

We train an HR teacher $T_H$ and an LR student $S_L$, each comprising an
encoder and a classifier head:
\begin{equation}
    z_i^H = e_H(x_i^H), \quad p_i^H = g_H(z_i^H),
\end{equation}
\begin{equation}
    z_i^L = e_L(x_i^L), \quad p_i^L = g_L(z_i^L).
\end{equation}
where $z_i^H$ and $z_i^L$ denote latent representations. The standard LR
baseline minimizes the cross-entropy objective,
\begin{equation}
    \mathcal{L}_{\mathrm{CE}} =
    \frac{1}{N}\sum_i \operatorname{CE}(p_i^L, y_i).
\end{equation}

RAST augments this LR classification objective with compact structural
supervision extracted from the HR teacher.\looseness=-1

\begin{figure}[t]
\centering
\includegraphics[width=\columnwidth]{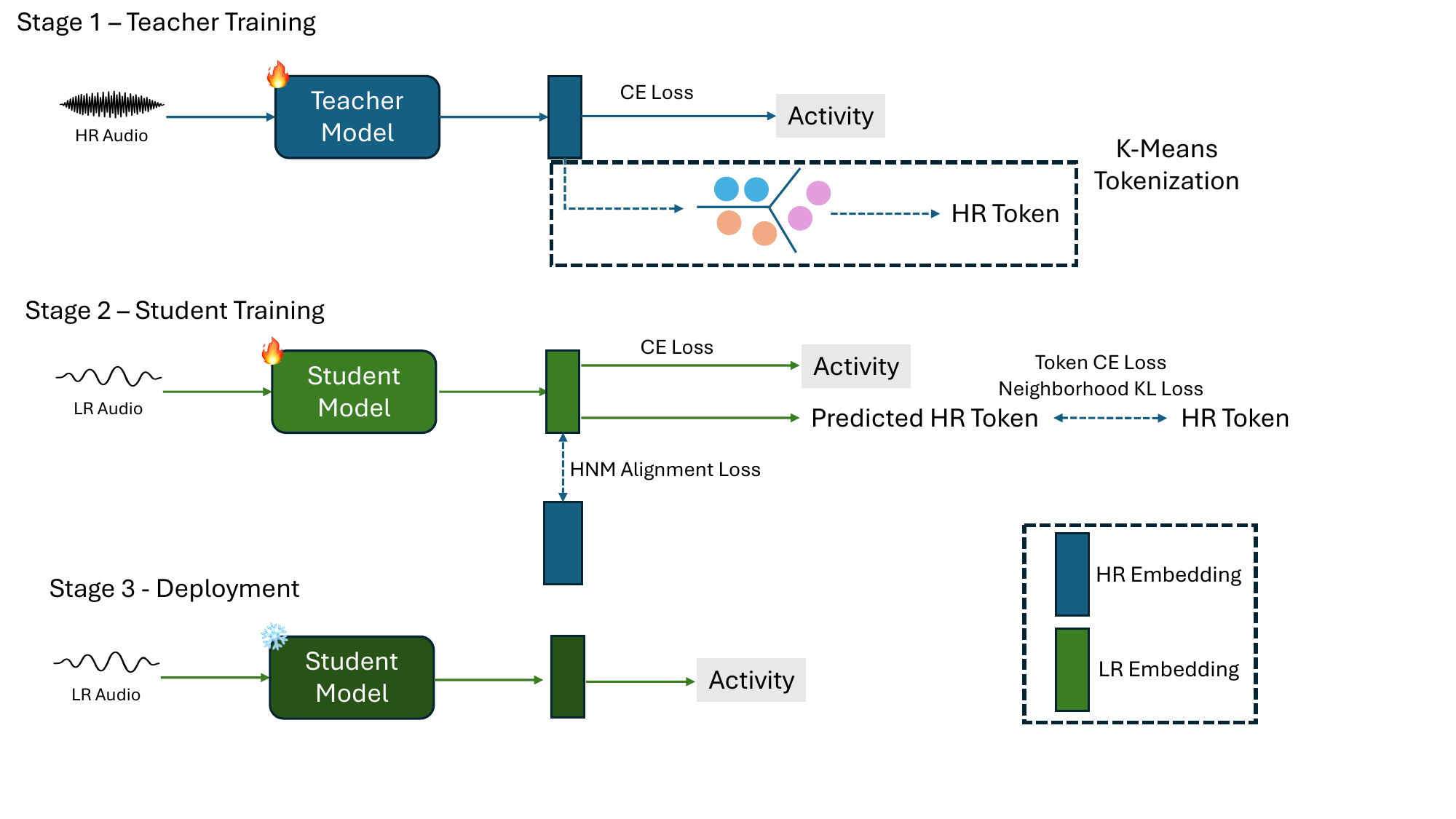}
\Description{Overall RAST training and deployment pipeline with 
three stages.}
\caption{RAST training and deployment pipeline. HR audio provides token- 
and neighborhood-level privileged supervision only during training, 
while deployment uses only the LR student and LR audio.\looseness=-1}
\label{fig:main_architecture}
\end{figure}

\subsection{Framework Overview}

RAST comprises three stages. In Stage 1, the HR teacher is trained for
activity recognition, and its intermediate representations are distilled
into two forms of compact privileged structure: discrete HR-derived
tokens and local teacher neighborhoods. In Stage 2, the LR student uses
these structures for initialization (warm start) and is subsequently
refined through local HR--LR representation alignment. In Stage 3, the
HR teacher and all auxiliary transfer targets are discarded. Deployment
therefore requires only the LR student and LR audio input.
Figure~\ref{fig:main_architecture} summarizes the privileged train-time 
transfer path and the LR-only inference path.

RAST transfers compact, structured HR knowledge rather than requiring the
LR student to reproduce the full HR embedding space. Because downsampling
removes acoustic evidence available to the teacher, direct HR--LR 
representation matching can impose partially unrecoverable targets. 
RAST therefore transfers bottlenecked HR-derived structure followed by 
local hard-negative alignment.

\subsection{Privileged Structure Extraction}

\textbf{HR-derived tokens.}
After training the HR teacher $T_H$, we extract representations from a 
fixed intermediate block and cluster the resulting feature space using 
$K$-means. Each training example receives a token label:\looseness=-1
\begin{equation}
    c_i = \arg\min_{k \in \{1,\ldots,K\}}
    \left\|\bar{z}_i^H - \mu_k\right\|_2^2,
\end{equation}
where $\bar{z}_i^H$ denotes the normalized HR teacher representation and 
$\mu_k$ the $k$th cluster centroid. These tokens provide a compact 
discretization of the HR representation space, exposing coarse teacher 
structure to the LR student. They serve only as auxiliary training targets 
and are discarded at inference.\looseness=-1

\textbf{Teacher neighborhoods.}
Although HR-derived tokens capture coarse regions of the teacher 
representation space, they do not preserve its local geometry. We therefore 
construct teacher-neighborhood distributions within each mini-batch. 
For samples $i$ and $j$, the HR pairwise similarity is:\looseness=-1
\begin{equation}
    s^H_{ij} = \operatorname{sim}(\bar{z}_i^H,\bar{z}_j^H),
\end{equation}
where $\operatorname{sim}$ denotes cosine similarity. The neighborhood
distribution induced by the HR teacher for anchor $i$ is:\looseness=-1
\begin{equation}
    q^H_{ij} =
    \frac{\exp(s^H_{ij}/\tau)}
    {\sum_{m\neq i}\exp(s^H_{im}/\tau)}, \quad j\neq i.
\end{equation}
where $\tau$ is the neighborhood temperature. The LR student induces an
analogous distribution $q^L_{ij}$ from $z^L$. We match these distributions
with:\looseness=-1
\begin{equation}
    \mathcal{L}_{\mathrm{nbr}} =
    \frac{1}{B}\sum_{i=1}^{B}
    \operatorname{KL}(q_i^H \| q_i^L),
\end{equation}
where $B$ is the mini-batch size.\looseness=-1

\subsection{Student Optimization and Alignment}

The LR student is first trained with label, token, and
teacher-neighborhood supervision:\looseness=-1
\begin{equation}
    \mathcal{L}_{\mathrm{warm}} =
    \mathcal{L}_{\mathrm{CE}}
    + \lambda_{\mathrm{tok}}\operatorname{CE}(r_i^L,c_i)
    + \beta \mathcal{L}_{\mathrm{nbr}},
\end{equation}
where $r_i^L$ is the LR token prediction head. This warm start shapes the LR
representation before direct HR--LR alignment.\looseness=-1

We then apply local hard-negative-mining (HNM) alignment between LR and HR
teacher embeddings. For anchor $i$, $(z_i^L,z_i^H)$ forms the positive pair, 
while $\mathcal{N}_k(i)$ contains the top-$k$ most similar non-matching 
HR examples selected as hard negatives in the mini-batch. We 
optimize:\looseness=-1
\begin{equation}
    \mathcal{L}_{\mathrm{HNM}} =
    -\frac{1}{B}\sum_i
    \log
    \frac{\exp(\operatorname{sim}(z_i^L,z_i^H)/\gamma)}
    {\sum_{j\in \mathcal{N}_k(i)}
    \exp(\operatorname{sim}(z_i^L,z_j^H)/\gamma)}.
\end{equation}
HNM focuses alignment on locally confusable examples rather than matching the
full HR manifold uniformly. The structured warm start further improves alignment 
by first exposing recoverable HR-derived tokens and neighborhood
structure to the LR student.\looseness=-1

\section{Experiments}

\subsection{Datasets and Sensing Setup}

We evaluate our proposed framework on wearable-based HAR datasets collected 
in home and semi-naturalistic environments: AudioIMU \cite{liang2022audioimu} 
and SAMoSA \cite{mollyn2022samosa}. AudioIMU contains 15 participants 
performing 23 activities of daily living, recorded with a Fossil Gen 4 
smartwatch at 22.05 kHz audio and 50 Hz IMU. SAMoSA contains 20 
participants recorded across diverse real-world environments using a 
Fossil Gen 5 smartwatch at 16 kHz audio and 50 Hz IMU. Both datasets 
provide synchronized acoustic and inertial streams; we use only 
audio.\looseness=-1

Audio is segmented into aligned 5-second windows. The HR branch uses 16 kHz
audio, while the LR branch simulates constrained sensing by reducing the
effective sampling rate to 1.25 kHz before computing the same log-mel
representation. HR and LR inputs are paired views of the same activity
window, with only LR audio available at test time.\looseness=-1

\subsection{Evaluation Protocol}

We use subject-independent evaluation to measure generalization to unseen
participants. SAMoSA uses leave-one-group-out validation over six participant
groups, while AudioIMU uses leave-one-participant-out validation over
participants 01--15. Results are reported as mean $\pm$ standard deviation over
five seeds. We report balanced accuracy (BA) and macro-F1; BA accounts for
uneven class frequencies, while macro-F1 measures per-class recognition without
allowing frequent classes to dominate.\looseness=-1

\subsection{Methods and Implementation}

All methods use the same PANNs-based audio encoder \cite{kong2020panns}, 
pretrained on AudioSet \cite{gemmeke2017audioset}, and identical 
subject-held-out folds. Baselines isolate different forms of HR-to-LR 
knowledge transfer: LR-only uses no privileged supervision; Knowledge 
distillation (KD) transfers teacher logits \cite{hinton2015distilling}; 
InfoNCE performs instance-level contrastive alignment 
\cite{oord2018cpc,tian2020crd}; Prototype NCE (ProtoNCE) aligns 
class-level prototypes inspired by prototypical and prototype-contrastive 
representation learning \cite{snell2017prototypical,li2021pcl}; and 
hard-negative mining (HNM) directly aligns local HR--LR representations using 
hard-negative selection \cite{robinson2021hardnegative}. RAST augments HNM 
with the proposed token and teacher-neighborhood warm start, evaluating 
whether compact HR structure transfers more effectively than direct teacher 
imitation under resolution mismatch.\looseness=-1

The HR token codebook uses $K=100$ clusters extracted from an intermediate 
teacher block. Unless otherwise stated, RAST uses $\lambda_{\mathrm{tok}}=1.0$, 
$\beta=0.1$, neighborhood temperature $\tau=0.2$, and HNM temperature 
$\gamma=0.2$. Experiments were conducted on a workstation with eight NVIDIA 
RTX A4500 GPUs, each with 20 GB of memory.\looseness=-1

\section{Results}

\subsection{Discussion on Overall Results}

Table~\ref{tab:main_results} compares RAST with LR-only training and direct
HR-to-LR transfer baselines. RAST consistently achieves the best LR inference
performance on both datasets. On SAMoSA, balanced accuracy improves from 
$0.6123$ to $0.6532$ and macro-F1 from $0.603$ to $0.6482$, corresponding to
relative gains of $6.7\%$ and $7.5\%$, respectively. On AudioIMU, balanced 
accuracy increases from $0.5157$ to $0.5474$ and macro-F1 from
$0.4933$ to $0.532$, yielding relative gains of $6.1\%$ and $7.8\%$ respectively.
Compared with KD, InfoNCE, ProtoNCE, and HNM, these gains indicate that HR 
supervision using teacher is more effective when first distilled into compact 
tokens and local neighborhood structure before localized HR--LR alignment.\looseness=-1

\subsection{Ablation Study}

\begin{table}[H]
\centering
\caption{Component-level performance of RAST on SAMoSA before and after HNM 
alignment. Balanced Accuracy (BA) and macro-F1 are reported.\looseness=-1}
\label{tab:ablation}
\begingroup
\renewcommand{\arraystretch}{0.98}
\begin{tabular*}{\columnwidth}{@{\extracolsep{\fill}}lcccc@{}}
\toprule
& \multicolumn{2}{c}{Warm start} & \multicolumn{2}{c}{After HNM} \\
\cmidrule(lr){2-3}\cmidrule(lr){4-5}
Init. & BA & F1 & BA & F1 \\
\midrule
CE          & $0.6123$ & $0.6030$ & $0.6218$          & $0.6093$          \\
Token       & $0.6218$ & $0.6105$ & $0.6443$          & $0.6240$          \\
Nbr.        & $0.6243$ & $0.6131$ & $0.6442$          & $0.6225$          \\
Token+Nbr.  & $0.6308$ & $0.6182$ & $\mathbf{0.6532}$ & $\mathbf{0.6482}$ \\
\bottomrule
\end{tabular*}
\endgroup
\end{table}

Table~\ref{tab:ablation} isolates how structured warm start supervision 
organizes the LR representation space before subsequent HNM alignment. With 
CE-only initialization, HNM improves balanced accuracy from $0.6123$ to 
$0.6218$. Adding HR-token supervision raises the warm-start score to $0.6218$ 
and the post-HNM result to $0.6443$, indicating that the token bottleneck 
provides coarse, activity-relevant anchors that facilitate subsequent local 
alignment.\looseness=-1

Teacher-neighborhood regularization further improves the alignment interface.
Although neighborhood supervision alone yields a moderate warm-start score, its
post-HNM performance is comparable to token supervision, suggesting that local
teacher geometry helps organize nearby examples before hard-negative alignment.
Combining tokens and neighborhoods yields the strongest warm start ($0.6308$ BA,
$0.6182$ macro-F1) and the final result ($0.6532$ BA, $0.6482$
macro-F1). These results support the RAST design: tokenization supervision 
provides compact HR-derived learnable anchors, neighborhood regularization 
structures their local geometry around these anchors, and HNM then aligns the 
resulting LR representation more effectively.\looseness=-1

\subsection{Embedding Structure Analysis}

\begin{figure}[H]
\centering
\includegraphics[width=\columnwidth]{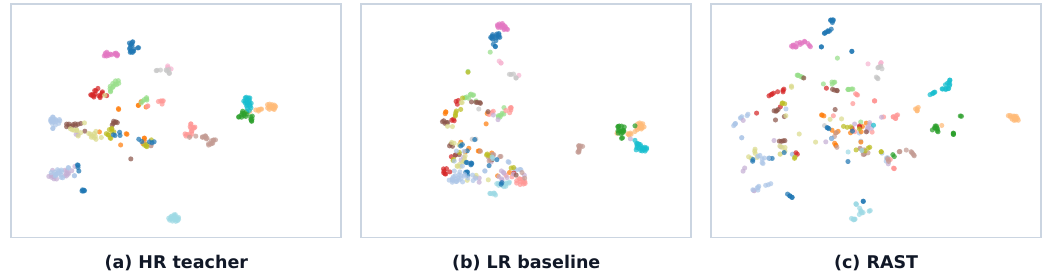}
\Description{Three t-SNE visualizations of intermediate embeddings for 
high-resolution, low-resolution, and RAST.}
\caption{t-SNE visualization of intermediate embeddings. RAST transfers HR-derived structure to the LR model, producing representations that are better aligned 
with activity structure than the LR baseline.\looseness=-1}
\label{fig:embedding_comparison}
\end{figure}

Figure~\ref{fig:embedding_comparison} illustrates the HR--LR representation gap
and the role of RAST's structured warm start. The HR teacher preserves more 
coherent local neighborhood geometry, with related activities forming 
better-organized regions. In contrast, the LR-only model produces a more 
entangled representation, particularly for activities that depend on fine 
acoustic cues attenuated by downsampling.\looseness=-1

RAST restructures the transfer problem before local alignment. Rather than 
directly matching the full HR embedding space, it first exposes the LR student 
to compact HR-derived tokens and teacher-neighborhood structure, providing coarse 
anchors and local geometric constraints for subsequent HNM. Consistent with the 
ablation results in Table~\ref{tab:ablation}, the resulting RAST embedding 
representation in Figure~\ref{fig:embedding_comparison} shows improved separation 
among locally confusable activities and aligns with the overall gains in 
Table~\ref{tab:main_results}.\looseness=-1

\section{Conclusion}

In this paper, we present RAST, a resolution-aware privileged structure transfer 
framework for low-resolution audio activity recognition. RAST uses HR audio only 
during training, compressing teacher representations into token- and 
neighborhood-level supervision before local HR--LR alignment. Across SAMoSA 
and AudioIMU datasets, RAST consistently outperforms LR-only training and 
direct teacher-student transfer under subject-independent evaluation.\looseness=-1

The HR--LR gap motivates methods that explicitly model resolution 
mismatch and LR observability. Future work includes student-aware teachers, 
adaptive privileged bottlenecks, and alignment objectives that preserve 
activity-relevant structure recoverable from LR audio, with broader 
evaluation of privacy and real-world deployment tradeoffs.\looseness=-1

\bibliographystyle{ACM-Reference-Format}
\bibliography{references}

\end{document}